\documentstyle[aps,epsfig,preprint]{revtex}

\tightenlines

\input{epsf}
\input{psfig.sty}

\begin{document}
\title{{\bf {\large Synchronous Behavior of Two Coupled Electronic Neurons \\
} }}
\author{R. D. Pinto$^1$, P. Varona$^{1,2}$, A. R. Volkovskii$^1$, A. Sz\"ucs$^1$,}
\author{Henry D. I. Abarbanel$^{1,3}$ and M. I. Rabinovich$^1$}
\address{$^1$Institute for Nonlinear Science,\\
University of California, San Diego, La Jolla, CA 92093-0402.\\
}
\address{$^2$GNB. ETS Ingenier\'{\i}a Inform\'{a}tica,\\
Universidad Aut\'{o}noma de Madrid, 28049 Madrid, SPAIN.}
\address{$^3$Department of Physics, and Marine Physical Laboratory,\\
Scripps Institution of Oceanography;\\
University of California, San Diego, La Jolla, CA 92093-0402}
\date{\today}
\maketitle

\begin{abstract}
We report on experimental studies of synchronization phenomena in a pair of
analog electronic neurons (ENs). The ENs were designed to reproduce the
observed membrane voltage oscillations of isolated biological neurons from
the stomatogastric ganglion of the California spiny lobster~{\em Panulirus
interruptus}. The ENs are simple analog circuits which integrate four
dimensional differential equations representing fast and slow subcellular
mechanisms that produce the characteristic regular/chaotic spiking-bursting
behavior of these cells. In this paper we study their dynamical behavior as
we couple them in the same configurations as we have done for their
counterpart biological neurons. The interconnections we use for these neural
oscillators are both direct electrical connections and excitatory and
inhibitory chemical connections: each realized by analog circuitry and
suggested by biological examples. We provide here quantitative evidence that
the ENs and the biological neurons behave similarly when coupled in the same
manner. They each display well defined bifurcations in their mutual
synchronization and regularization. We report briefly on an experiment on
coupled biological neurons and four dimensional ENs which provides further
ground for testing the validity of our numerical and electronic models of
individual neural behavior. Our experiments as a whole present interesting
new examples of regularization and synchronization in coupled nonlinear
oscillators.


\end{abstract}

\section{Introduction}

\bigskip \bigskip Synchronization of nonlinear oscillators is widely studied
in physical and biological systems~\cite{glass,neural} for underlying
interests ranging from novel communications strategies~\cite{lasers1,lasers2}
to understanding how large and small neural assemblies efficiently and
sensitively achieve desired functional goals~\cite{neuroscience}. The
analysis of biological systems may, beyond their intrinsic interest, often
provide physicists with novel dynamical systems possessing interesting properties in
their component oscillators or in the nature of the interconnections.

We have presented our analysis of the experimental synchronization of two
biological neurons as the electrical coupling between them is changed in
sign and magnitude~\cite{prl}. Subsequent to that analysis we have developed
computer simulations of the dynamics of the neurons which are based on
conductance based Hodgkin-Huxley (HH)~\cite{HH} neuron models. These
numerical simulations quantitatively reproduced the observations in the
laboratory~\cite{prl,martin,physicaa,elecpaper}.

The study of isolated neurons from the stomatogastric ganglion (STG) of the
California spiny lobster~{\em Panulirus interruptus} using tools of
nonlinear time series analysis~\cite{book,kantz} shows that the number of
active degrees of freedom in their membrane potential oscillations typically
ranges from three to five~\cite{neural}. The appearance of low dimensional
dynamics in this biological system led us to develop models of its action
potential activity which are substantially simpler than the HH models we and
others~\cite{martin} have used to describe these systems. We adopted the
framework established by Hindmarsh and Rose (HR)~\cite{HR} in which the
complicated current voltage relationships of the conductance based models
are replaced by polynomials in the dynamical variables, and the coefficients
in the polynomials are estimated by analyzing the observed current/voltage
curves for the neurons under study.

Building on biological experiments and on numerical analysis of models for
the oscillations of isolated neurons, we have constructed low dimensional
analog electronic neurons (ENs) whose properties are designed to emulate the
membrane voltage characteristics of the individual neurons. Using these
simple, low dimensional ENs we report here their synchronization and
regularization properties, first when they are coupled electrically as the
sign and magnitude of the coupling is varied, and then when they are coupled
by excitatory and inhibitory chemical synapses. We have also studied the
behavior of an hybrid system, i.e., one EN and one biological neuron coupled
electrically. As our models were developed on data acquired
from biological neurons in synaptic isolation, the results we present here
on pairs of interacting ENs and hybrid systems serve to provide further
confirmation of the properties of those model neurons, numerical and analog.

\section{The Electronic Neuron Model}

\label{sectioneqs}

We have studied and built three dimensional and four dimensional models of
HR type having the form 
\begin{eqnarray}
\frac{d x(t)}{dt} &=& a y(t) + b x^2(t) - c x^3(t) - dz(t) + I  \nonumber \\
\frac{d y(t)}{dt}&=& e - f x^2(t) - y(t) - g w(t)  \nonumber \\
\frac{d z(t)}{dt} &=& \mu (-z(t) + S(x(t) + h))  \nonumber \\
\frac{d w(t)}{dt} &=& \nu (-k w(t) + r(y(t)+l)),  \label{eq:HRmodel}
\end{eqnarray}
where $a,b,c,d,I,e,f,g,\mu,S,h,\nu,k,r$ and $l$ are the constants which
embody the underlying current and conductance based dynamics in this
polynomial representation of the neural dynamics. $x(t)$ is the membrane
voltage in the model, $y(t)$ represents a ``fast'' current in the ion
dynamics, and we choose $\mu \ll 1$, so $z(t)$ is a ``slow'' current. Taken
alone the first three equations of the model can reproduce several modes of
spiking-bursting activity observed in STG cells. The first three equations
were used in analog realization for our earlier experiments with 3D ENs~\cite
{Szu00}

The equation for $w(t)$ represents an even slower dynamical process ($\nu <
\mu \ll 1$), and it is included because a slow process such as the calcium
exchange between intracellular stores and the cytoplasm was found to be
required in Hodgkin-Huxley modeling~\cite{martin} to fully reproduce the
observed chaotic oscillations of STG neurons. Both the three dimensional and
four dimensional models have regions of chaotic behavior, but the four
dimensional model has much larger regions in parameter space where chaos
occurs, presumably for many of the same reasons the calcium dynamics gives
rise to chaos in HH modeling. The calcium dynamics is an additional degree
of freedom with a time constant three times slower than the characteristic
bursting times.

In our analog circuit realization of the EN model we used $a = 1, b = 3, c =
1, d = 0.99, I = 3.024, e = 1.01, f = 5.0128, g = 0.0278, \mu =0.00215, S =
3.966, h = 1.605, \nu = 0.0009, k = 0.9573, r = 3.0$ and $l = 1.619$. The
implementation of these constants in analog circuits always has about 5\%
tolerance in the components. The main parameters we used in controlling the
modes of spiking and bursting activity of the model are the DC external
current $I$ and the time constants $\mu$ and $\nu$ of the slow variables.

\begin{figure}[ht!]
\centerline{
\epsfig{file=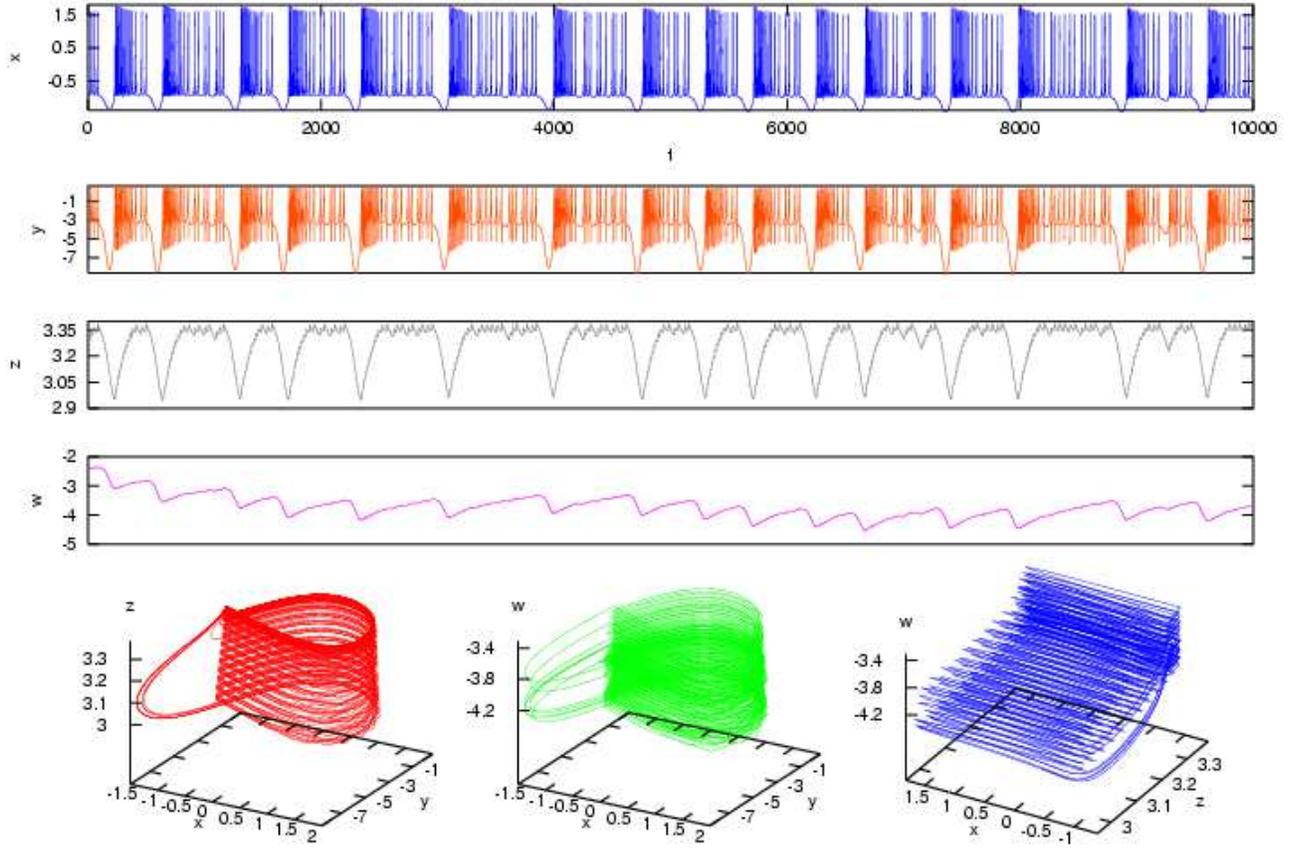,width=12cm,angle=-90}
}
\caption{\small{Time series of the dynamical variables $x(t), y(t), z(t), w(t)$ of our four dimensional HR model, Equation(\ref{eq:HRmodel}), and various
three-dimensional projections $(x(t),y(t),z(t))$, $(x(t),y(t),w(t))$ and $(x(t),z(t),w(t))$ of the four dimensional
phase space orbits.. }}
\label{4dphases}
\end{figure}

Figure~\ref{4dphases} shows a chaotic time series of the four variables
using the parameters above. Note how $w$ modulates the length of the bursts
in $x$. Each local minimum in the global oscillations of $w$ coincides with
a short burst period. The complexity achieved by the addition of $w$ can be
observed in the projections of $(x,y,z,w)$ space to various
three-dimensional spaces, $(x,y,z),(x,y,w)$ and $(x,z,w)$ respectively, as
shown in Figure~\ref{4dphases}.

Table 1 presents the Lyapunov exponents $\lambda_i$ calculated from the
vector field~\cite{parker} of Equation(~\ref{eq:HRmodel}) for the 3D and 4D
ENs. A positive Lyapunov exponent is present in both models, indicating
conclusively that they are oscillating chaotically. From this spectrum of
Lyapunov exponents, we can evaluate the Lyapunov dimension $D_L$ which is an
estimate of the fractal dimension of the strange attractor for the ENs~\cite
{book}. The Lyapunov dimension is defined by finding that number of Lyapunov
exponents $\lambda_i$ satisfying 
\begin{equation}
\sum_{i=1}^N\lambda_i > 0\;\;;\mbox{while}\;\;\sum_{i=1}^{N+1}\lambda_i < 0.
\end{equation}
Then $D_L$ is defined as 
\begin{equation}
D_{L}=N+\frac{\sum_{i=1}^{N}\lambda_i}{|\lambda_{N+1}|}.
\end{equation}
$D_L$ for each EN is displayed in the last column Table 1.

\begin{center}
\begin{tabular}[t]{|c|c|c|c|c|c|}
\hline
Model: & $\lambda_1$ & $\lambda_2$ & $\lambda_3$ & $\lambda_4$ & $D_{L}$ \\ 
\hline
3D & 0.010 & 0.000 & -7.752 & $-$ & 2.001 \\ \hline
4D & 0.004 & 0.000 & -0.001 & -8.034 & 3.000 \\ \hline
\end{tabular}
\end{center}

Table 1: {\small Lyapunov exponents $\lambda_i$ and Lyapunov dimension $D_L$
calculated from the vector field (Equation(~\ref{eq:HRmodel})) for the 3D
and 4D electronic neuron models. As a reminder to the reader: the sum of all
Lyapunov exponents must be negative, and this is so for our results. Also,
one Lyapunov exponent must be 0 as we are dealing with a differential
equation.}

\section{Analog Implementation of the ENs}

\label{electronicneurons}

We designed and built an analog electronic circuit which integrates Equation
(\ref{eq:HRmodel}). We chose to build an analog device instead of using
numerical integration of the mathematical model on the CPU of a PC or on a
DSP board because digital integration of these equations is a slow procedure
associated with the three different time scales in the model. Furthermore, a
digital version of an EN requires digital to analog and analog to digital
converters to connect the model to biological cells. Analog circuits are
small, simple and inexpensive devices; it is easy to connect them to a
biological cell, as we discuss below (see also~\cite{Szu00}). In a practical
sense nearly an unlimited number of them can work together in real-time
experiments. Finally, looking ahead to the construction of real-time
networks of large number of these neurons, analog implementation is a
necessity.

\begin{figure}[ht!]
\centerline{
\epsfig{file=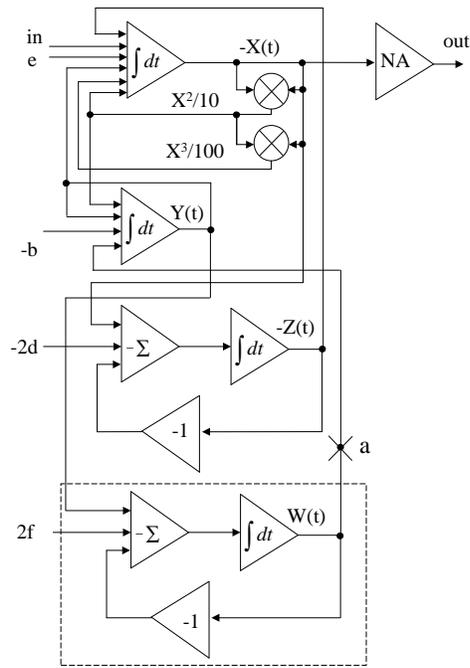,width=8cm}
}
\caption{\small{Block diagram for the four dimensional
HR neuron used in our experiments. These neurons were designed to
replicate the behavior of individual, isolated neurons from the lobster STG. In
our experiments they were coupled electrically as well as via an electronic
implementation of inhibitory and excitatory chemical synapses.
}}
\label{4dcircuit}
\end{figure}

The block diagram of the analog circuit we use to represent the three
dimensional and the four dimensional ENs is shown in Figure (\ref
{4dcircuit}). It includes four integrators indicated by $\int dt$, two
multipliers, two adders, and two inverters. We used off-the-shelf
general purpose operational amplifiers, National Instruments Model
TL082, to build the integrators, adder and inverter and used Analog
Devices Model AD633 as analog multipliers. The integrator at the top
of the diagram receives all components of $\frac{dx(t)}{dt}$,
e.g. $ay(t)$, $bx(t)^3$, etc. It has an additional input (called $in$)
which can be used for connections with other circuits or neurons. The
integrators invert the sign of their input, so the output signal will
be $-x(t)$ multiplied by a time constant $\tau$ chosen to make the EN
oscillate on the same time scale as the biological neurons. The signal
$-x(t)$ is used to generate the nonlinear functions $x^2(t)$ and
$x^3(t)$ and these values go to the second and third
integrators. Similarly, the other integrators generate voltages
proportional to $y(t)$, $-z(t)$, and $w(t)$. A renormalization was
made in the rest of the time constants in the circuit to make
$\tau=1$. Note that this rescaling is responsible for the different
amplitudes in the numerical (Figure~\ref{4dphases}) and analog
(Figures~\ref{free},~\ref{posel},~\ref{negel},~\ref{exchem},~\ref{inchem})
experiments.

This circuit design allows us to easily switch from a three dimensional to a
four dimensional model of the neuron. We can connect or disconnect one wire,
indicated as point 'a' in Figure~\ref{4dcircuit}, to enable or disable the
circuit block shown in the rectangle with a dashed outline. In Equation(\ref
{eq:HRmodel}) this corresponds to setting $g=0$ in the $\frac{dy(t)}{dt}$
equation.

The block indicated as NA in Figure~\ref{4dcircuit} is an adjustable
nonlinear amplifier. We use it to rescale and change the shape of the
output signal $x(t)$. It can shrink or stretch different parts of the
waveform, change the amplitude and move the trace as a whole up or
down. This shape adjustment is particularly important in experiments
with groups of biological and electronic neurons interconnected with
each other. Living neurons, even taken from the same biological
structure, may generate considerably different waveforms. The relative
size of spikes and the interburst hyperpolarization is variable from
cell to cell. In our circuits we can precisely adjust the waveform of
the EN to be very close to that of each biological neuron in our
experiments.

Another reason to use circuits with variable waveforms is that it opens up
the possibility of studying how the action potential waveforms affect the
interactions among the neurons, electronic and biological. Indeed, the
ability to vary the details of the waveforms provides an interesting handle
on design of biometric circuitry for a variety of applications.

\section{Synaptic Connections Between ENs}

In living nervous systems one finds three general types of synaptic
connections among neurons~\cite{shep}: ohmic electrical connections (also
called gap junctions) and two types of chemical connections, excitatory and
inhibitory. For our studies of the interconnections among ENs and among ENs
and biological neurons~\cite{Szu00}, we built electronic circuits to emulate
excitatory and inhibitory synaptic connections as well as the ohmic
electrical connections. The STG neural circuits are dominated by inhibitory
interconnections and by ohmic electrical connections. We now describe how we
implemented each, and then we turn to the results of our synchronization
experiments with these network connections.

\subsection{Implementation of the Electrical Synapses}

We implemented an electrical synapse~\cite{Sha92} between the ENs by
injecting into one of the neurons ($EN_1$) a current proportional to the
voltage difference between the two membrane potentials of the ENs and into
the other neuron ($EN_2$) injecting the same current but with the opposite
strength. The current into $EN_1$ is

\begin{center}
$I_{1}(t)=\frac{G_E}{470k\Omega }(x_{1}(t)-x_{2}(t));$ \\[0pt]
while $I_{2}(t)=-I_{1}(t).$
\end{center}

We chose the dimensionless synaptic strength $G_{E}$ in the range $G_E \in %
\left[ -1,1\right]. $ Over this range we observed the effects of positive
and negative electrical coupling on the spiking and bursting behavior of the
ENs. We recorded the electrical voltage signals corresponding to the
membrane potentials of the ENs using an analog to digital converter with a
sampling rate of 5 kHz. For each value of $G_{E}$ we waited at least 40
seconds to avoid transient dynamics and then recorded a data series 20
seconds long.

Natural biological networks do not have negative conductance electrical
coupling. Using an active device placed between the neurons we implemented
negative electrical couplings in our experiments on two electrically coupled
biological neurons as reported in~\cite{prl}. To compare the results of our
work there with the properties of coupled ENs, we use negative coupling here
as well.

\subsection{Implementation of the Chemical Synapses}

We first implemented mutual chemical synapses between the two ENs
using analog circuitry. Here we report on results
obtained by using a software implementation of the chemical synapses
which allows to investigate the role of the synaptic time constant
$\tau _{s}$. In the analog circuit implementation of the chemical synapses we
need to replace a capacitor every time we want to change the time
constant, but in the software version this time
constant is just a parameter, so it is easier to study the role of these
time constant in the software version. In this paper the time constant is fixed, and our observations on the role of a changing time constant will be reported in
another paper. The results using a software version of the chemical synapse,
and the results using our hardware version were identical.

We used the nonlinear amplifiers to reshape the signals
corresponding to the membrane potential of the ENs in such a way that
the new signals had amplitudes, spike/burst ratios, and voltage
offsets close to the the signals generated by living neurons. With
these reshaped signals we used new dynamic clamp
software~\cite{reynaldo} to generate in real time the currents
corresponding to the graded chemical synapses as described the by
first-order kinetics

\begin{equation}
I_{C}=213G_{c}S(t)(x_{\mbox{rev}}-x_{\mbox{post}}),
\end{equation}
and 
\begin{equation}
(1-S_{\infty })\tau _{s}\frac{dS(t)}{dt}=(S_{\infty }-S(t))
\label{chemsynapse}
\end{equation}
where 
\begin{equation}
S_{\infty }(x_{\mbox{pre}})= \tanh [\frac{x_{\mbox{pre}}-x_{\mbox{th}}}{x_{%
\mbox{slope}}}],
\end{equation}
when $x_{\mbox{pre}}>x_{\mbox{th}}$. Otherwise $S_{\infty }(x_{\mbox{pre}})=0
$.

$G_{c}$ is the maximal synaptic conductance, $S(t)$, the instantaneous
synaptic activation, $S_{\infty }$, the steady-state synaptic activation, $%
x_{\mbox{rev}}$, the synaptic reversal potential, and $x_{\mbox{pre}}$ and $%
x_{\mbox{post}}$ are the presynaptic and postsynaptic voltages respectively. 
$\tau _{s}$ is the synaptic time constant, $x_{\mbox{th}}$,
the synaptic threshold voltage, and $x_{\mbox{slope}}$, the synaptic slope
voltage.

The synaptic reversal potentials were selected so that the currents injected
into the postsynaptic ENs were always negative for inhibitory synapses and
positive for excitatory synapses, emulating the biological synapses~\cite
{shep}. The synaptic threshold voltages were set in the middle of the
amplitude of the bursts, and the synaptic slope voltage was adjusted to make
the output of the hyperbolic tangent slightly saturated at the spikes. In
our experiments $G_{c}$ was varied as we collected different data sets. We
used standard values for the other parameters in the dynamic clamp program: $%
x_{\mbox{rev}}=-80mV$ (inhibitory synapses) or $x_{\mbox{rev}}=-20mV$
(excitatory synapses); $\tau _{s}=10ms$; $x_{\mbox{th}}=-50mV$; and $x_{%
\mbox{slope}}=10mV$. As before we waited for at least 40 seconds after
connecting the ENs with the chemical synapses before starting the recording
of the 20 seconds of data from the membrane potential of the ENs.

\section{Experiments}

To analyze the degree of synchronization of slow bursts between two coupled
neurons (electrically or chemically) we proceed in the same manner as we
used for our experiments on synchronized living neurons~\cite{prl}. This was
based on a method developed for the experimental studies of synchronization
of chaotic oscillations in electronic circuits~\cite{Afr86,heagy,rulkov}. We
used an overlap-add method of FIR filtering with a Hamming window, and used an FFT
and a cutoff frequency of 5Hz to suppress the spikes, obtaining the filtered
data series $x^{f}_{i}(t)$; $i=1,2$. The synchronization of the ENs is
quantified by calculating the difference $%
x^{f}_d(t)=x^{f}_{1}(t)-x^{f}_{2}(t)$, and studying the normalized standard
deviation $\sigma _{N}=\sigma_{x^{f}_d}/\sigma _{x^{f}_{1}}$ and the
normalized maximal deviation $\Delta _{N}=\left| x^{f}_d\right|^{\max
}/(x^{f,\max}_{1}-x^{f,\min}_{1})$ as a function of $G_{E}$ for the
electrical coupling or as a function of $G_{c}$ for the chemical coupling.
For notational convenience, we indicate excitatory couplings with values of $%
G_c > 0$ and inhibitory couplings with values of $G_c < 0$.

\subsection{Isolated Neurons:}

\begin{figure}[ht!]
\centerline{
\epsfig{file=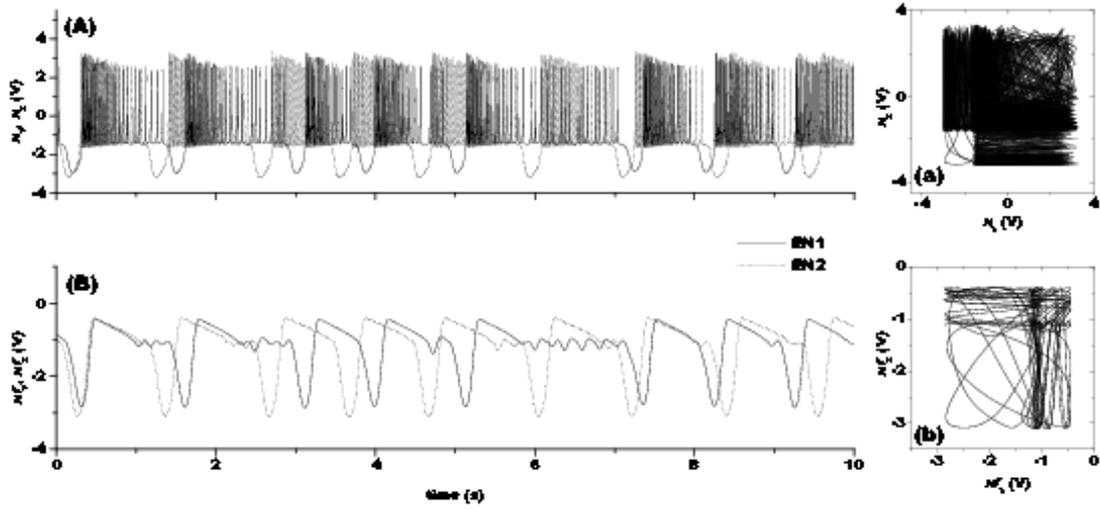,width=15cm}
}
\caption{\small{Regimes of oscillations in two uncoupled electronic neurons (ENs). (A) Time
series of the membrane voltages $x_1(t),x_2(t)$ for the {\bf uncoupled} ENs. (a) Phase space portraits of $x_{2}(t)
$ vs. $x_{1}(t)$. (B) Membrane potentials after 20Hz low-pass filtering to
emphasize the bursting behavior. (b) Phase space portraits of filtered membrane
potentials $x^{f}_{2}(t)$ vs. $x^{f}_{1}(t)$.}}
\label{free}
\end{figure}

The parameters of the isolated neurons were set in the chaotic
spiking-bursting regime. An example of the behavior of an isolated EN is
shown in Figure~\ref{free}. Note that the scale for $x$ is double that of
the numerical simulations shown in Figure~\ref{4dphases} because of the
rescaling time constant in the analog integrator (see section~\ref
{electronicneurons}). The relative behavior of the spikes and slow
oscillations can be seen in the plots of $x_{2}$ vs. $x_{1}$ (Figure~\ref
{free}A) and $x^{f}_{2}$ vs. $x^{f}_{1}$ (Figure~\ref{free}B), respectively.

\subsection{Electrical Coupling Between Two ENs}

\begin{figure}[ht!]
\centerline{
\epsfig{file=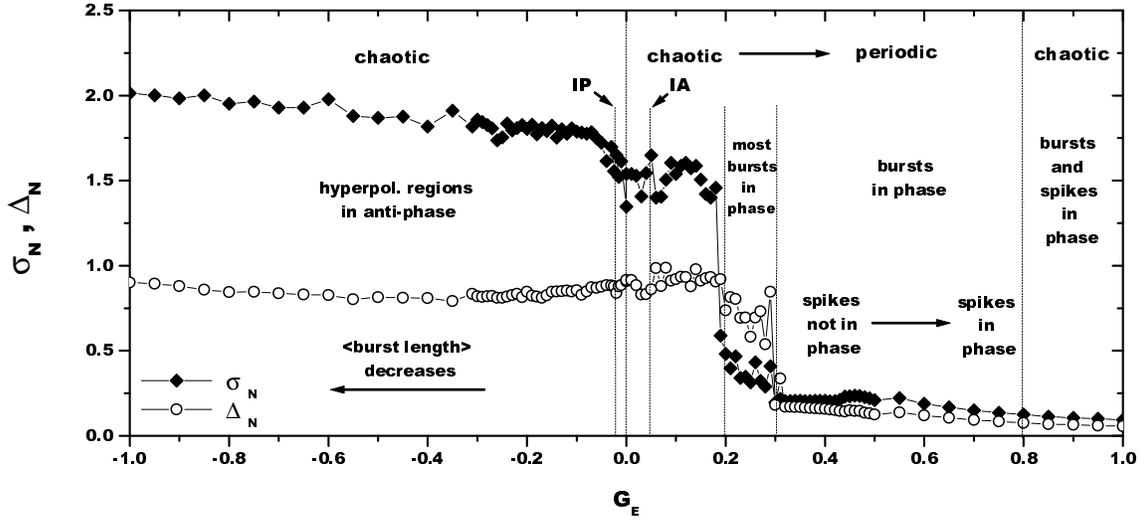,width=16cm}
}
\caption{\small{Normalized standard deviation $\sigma _{N}(G_E)$ and normalized maximal deviation 
$\Delta _{N}(G_E)$ computed after 20Hz low-pass filtering the membrane
potential of two electrically coupled ENs for different values of the
synaptic conductance $G_{E}$. For small positive electrical coupling $(G_{E}>\approx 0)$
the behavior of the ENs is nearly independent and chaotic. Intermittent
out-of-phase bursting activity was observed in the range labeled as IA
in the diagram for the two coupled ENs with $G_{E}=0.05$. A transition from
chaotic to periodic behavior occurs for $0.2<G_{E}<0.3$ in which the chaotic
oscillations are replaced by synchronized bursting activity. In the interval 
$0.3<G_{E}<0.8$ smooth synchronization of the spikes occurs, and for $G_{E}>0.8$
the oscillations become chaotic but now with both bursts and spikes
synchronized. For negative coupling the oscillations are always chaotic and
the hyperpolarizing regions are in out-of-phase. There is a decrease in the
average burst length as the negative coupling become stronger. For a very
small negative coupling an intermittent behavior in which the
ENs showed very long and simultaneous bursts  was observed. This is the
region labeled as IP for $G_{E}=-0.02$.}}
\label{mapel}
\end{figure}

We began with electrical coupling between two four dimensional analog
circuit models implementing Equation(\ref{eq:HRmodel}). We varied only $G_{E}
$ keeping all other parameters fixed. A convenient representation of the
range of behavior we observed is presented in Figure~\ref{mapel}. Here,
overlying values of $\sigma_N(G_E)$ and $\Delta_N(G_E)$ we give a verbal
description of the quantitative behavior of time series in each regime. To
illustrate the phenomena seen in each regime $G_E \in [-1,1]$ of Figure~\ref
{mapel} we show examples of the time series for the membrane potentials $%
x_1(t),x_2(t)$ of the two neurons in Figures~\ref{posel} and ~\ref{negel}.

\subsubsection{Results for $0 \le G_E \le 1$}

\begin{figure}[ht!]
\centerline{
\epsfig{file=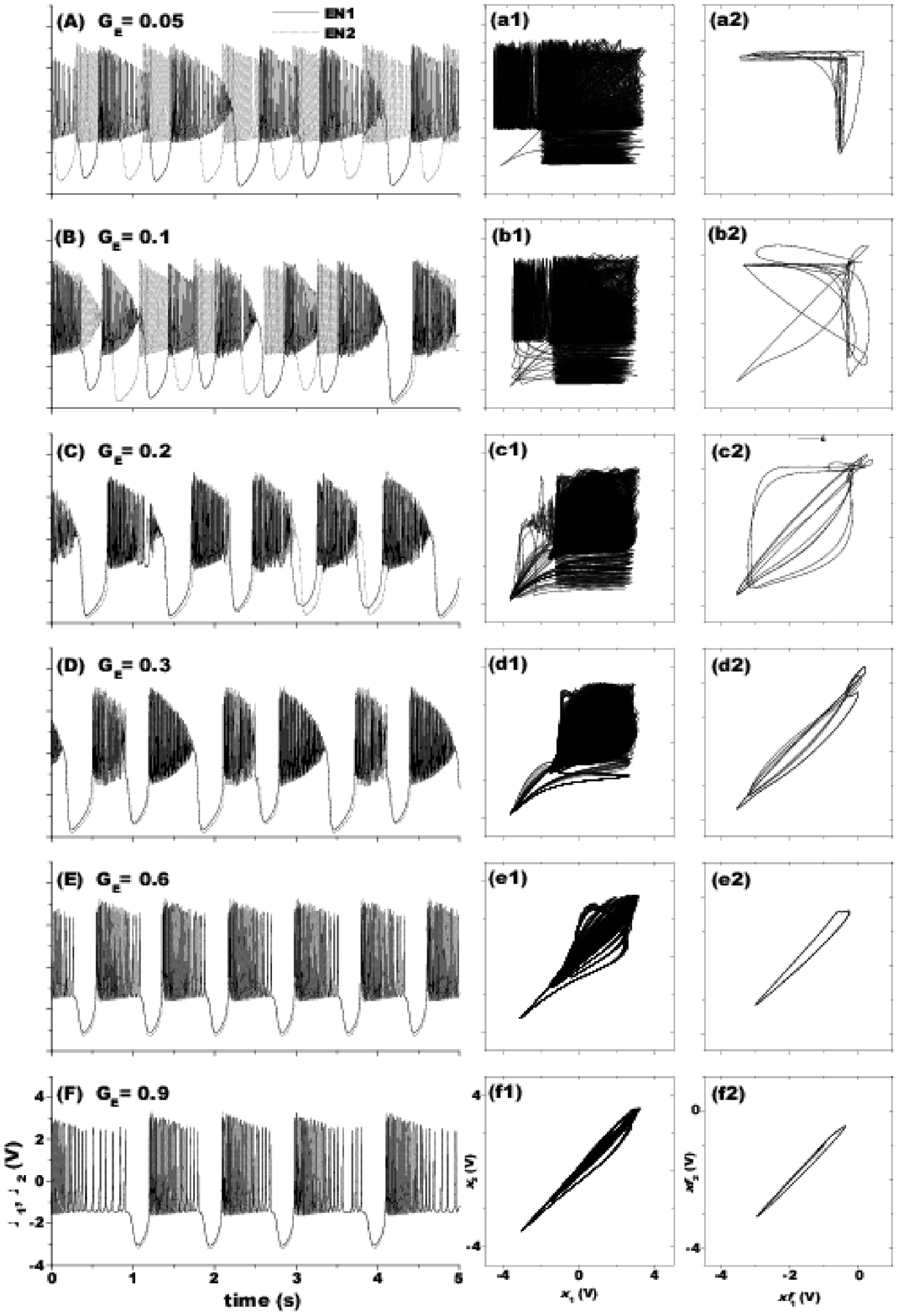,width=16cm}
}
\caption{\small{Positive electrical coupling of two chaotic ENs. Characteristic time series
of the membrane potentials $x_1(t),x_2(t)$ (figure labels A through F) as we vary $G_E$.
Phase portrait $x_{2}(t)$ vs. $x_{1}(t)$ (figure labels a1 through f1). Phase protraits after 20Hz
low-pass filtering $x^{f}_{2}(t)$ vs. $x^{f}_{1}(t)$ (figure labels a2 through f2). 
(A) $G_{E}=0.05$--intermittent out-of-phase bursting activity. (B) $G_{E}=0.1$--nearly independent 
chaotic spiking-bursting pattern. (C) $G_{E}=0.2$--chaotic oscillations with most bursts synchronized. (D) $G_{E}=0.3$--periodic oscillations with partial synchronization of the ENs, the spikes
are not synchronized. (E) $G_{E}=0.6$---periodic oscillations with the
complete synchronization of the ENs. (F) $G_{E}=0.9$--chaotic but
completely synchronized oscillations.}}
\label{posel}
\end{figure}

\begin{figure}[ht!]
\centerline{
\epsfig{file=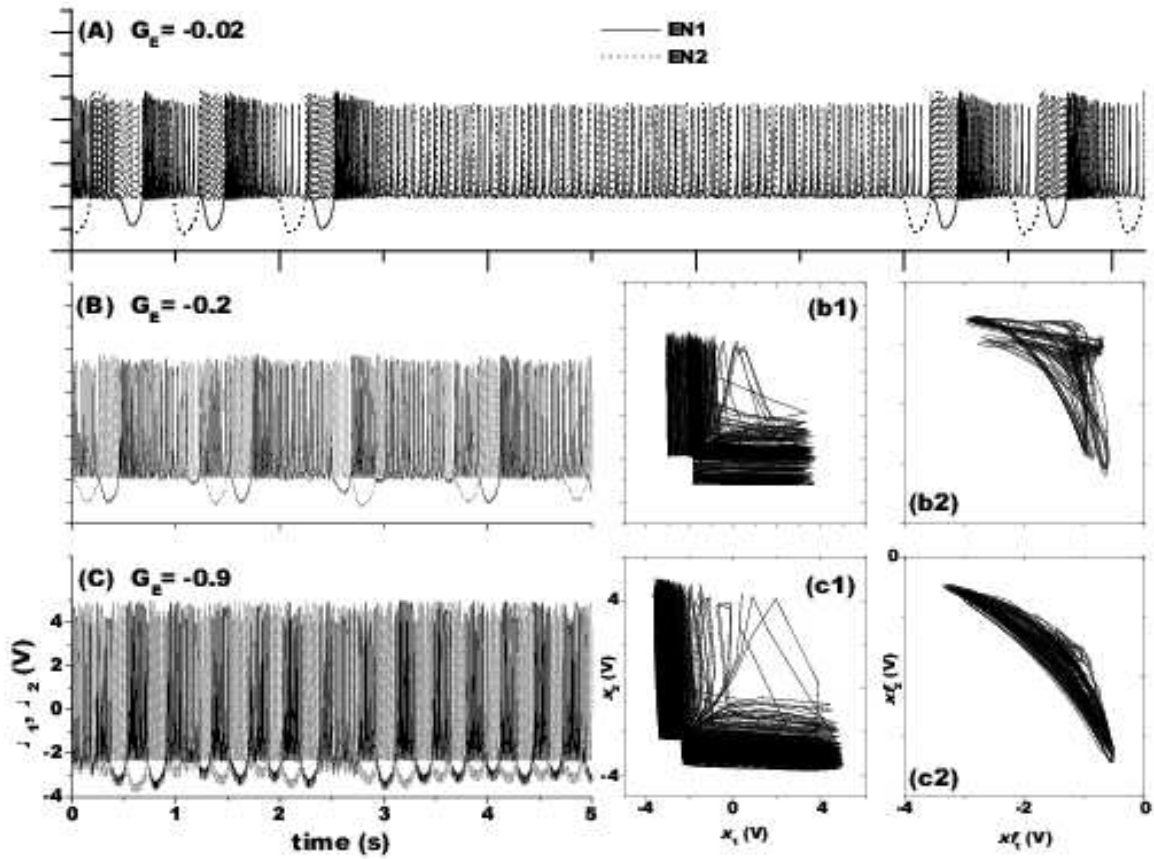,width=16cm}
}
\caption{\small{Negative electrical coupling of two chaotic ENs. Characteristic time series
of the membrane potentials $x_1(t),x_2(t)$ (figure labels A through C) as we vary $G_E$. Phase portraits of $x_{2}(t)$ vs. $x_{1}(t)$ (figure labels b1 and c1). Phase portraits after 20Hz 
low-pass filtering $x^{f}_{2}(t)$ vs. $x^{f}_{1}(t)$ (figure labels b2 and c2). 
(A) $G_{E}=-0.02$--intermittent simultaneous long bursts in the two ENs. (B) $G_{E}=-0.2$--chaotic 
out-of-phase spiking-bursting behavior. (C) $G_{E}=-0.9$--fast chaotic out-of-phase spiking-bursting behavior. The time scale used in the time series plot is the same for all examples.}}
\label{negel}
\end{figure}

\begin{itemize}
\item  When $G_{E}\approx 0.0$ the two neurons are uncoupled and display
independent chaotic oscillations as shown in Figure~\ref{free}.

\item  For small, positive coupling $0.0<G_{E\text{ }}<0.2$, regions of
nearly independent chaotic spiking-bursting activity are observed as well as
some regions of synchronized bursting activity as shown in Figure~\ref{posel}%
(B) where we set $G_{E}=0.1$. There is a small range of $G_{E\text{ }}$ $%
(G_{E}\approx 0.05)$ in which intermittent anti-phase bursting behavior can
be found. The burst length in this case is kept nearly regular from burst to
burst as shown in Figure~\ref{posel}(A).

\item  For $0.2 \leq G_{E} < 0.3$ the behavior is still chaotic for the two
neurons but most of the bursts are synchronized as shown in Figure~\ref
{posel}(C) where we set $G_{E}=0.2$.

\item  From $0.3 \leq G_{E}<0.8$ the bursting activity becomes regular
going from a region in which there is partial synchronization (spikes not
synchronized), as shown in Figure~\ref{posel}(D) where we set $G_{E}=0.3$,
to a region of total synchronization (bursts and spikes synchronized), shown
in figure~\ref{posel}(E) where we set $G_{E}=0.6$.

\item  From $0.8 \leq G_{E}<1.0$ there is total synchronization in the
spiking-bursting activity, and the oscillations are chaotic as shown in
figure~\ref{posel}(F) where we set $G_{E}=0.9$.

Results for $-1 \le G_E \le 0$

\item For negative coupling $G_{E\text{ }}<0$, the oscillations are
predominantly chaotic and the hyperpolarizing regions, where the
membrane voltage is quite negative, of the signals are all in
anti-phase. The average burst length decreases as the coupling becomes
stronger as shown in Figure~\ref{negel}.  For a small range of $G_{E}$
$(G_{E}\approx -0.02)$ very long bursts were observed as shown in
Figure~\ref{negel}(A).
\end{itemize}

\subsubsection{Comparison of Coupled ENs \\
with Electrically Coupled Biological Neurons}

$\sigma_N(G_E)$ and $\Delta_N(G_E)$ provide quantitative measures of the
synchronization between two ENs. In our report on the experimental work~\cite
{prl} with two biological cells, the results for $\sigma_N(G_E)$ and $%
\Delta_N(G_E)$ can be seen in Figure 5 of that paper. Note that, as in the
case of coupled biological neurons, we have here a bifurcation between
positive and negative electrical coupling. In the experimental work on
electrically coupled biological neurons a value for the external coupling $%
g_a \approx -200\,$nS serves to null out the natural coupling of about that
amount, so the figures here and in the earlier paper are to be compared by
sliding $G_{E}=0$ here to $g_a \approx -200\,$nS there. Both in the
biological and electronic experiments, the sharp phase transition from very
small $\sigma_N, \Delta_N$ for positive coupling to large, nearly constant
values is associated with the rather rapid change from nearly and then fully
synchronous behavior for positive couplings to out-of-phase oscillations for
negative couplings.

The $\sigma_N(G_E)$ and $\Delta_N(G_E)$ curves in the paper on coupled
biological neurons~\cite{prl} shows far fewer points and consequently less
detail that our curves for coupled 4D ENs. Clearly this is because of the
resolution in the biological experiments and the difficulty in performing
experiments at such closely chosen values of $G_E$. At this time the details
of behavior revealed in the present experiments on ENs have not been
verified in the biological setting. One should view our Figure~\ref{mapel}
and Figure 5 of~\cite{prl} as in excellent qualitative agreement.

\subsection{Chemical Synapses Between Two ENs}

\begin{figure}[ht!]
\centerline{
\epsfig{file=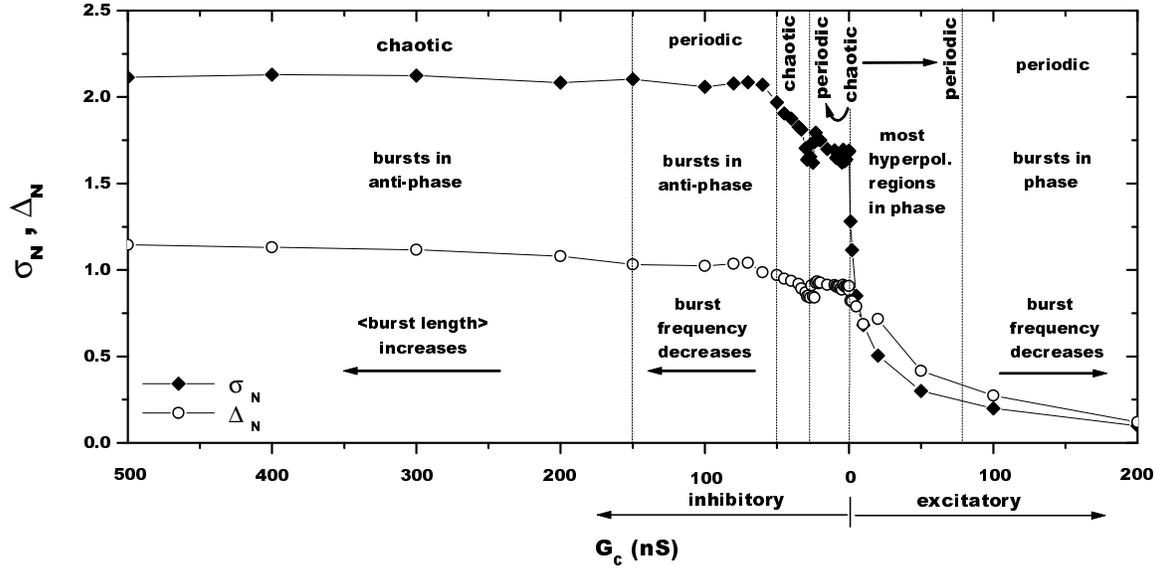,width=16cm}
}
\caption{\small{Normalized standard deviation $\sigma _{N}(G_c)$ and
normalized maximal deviation $\Delta _{N}(G_c)$ computed after 20Hz
low-pass filtering in the membrane potential of two ENs coupled with
identical chemical synapses for different values of the synaptic
conductance $G_{c}$ for both excitatory ($G_c > 0$) and inhibitory ($G_c<0$)
connections. For very small excitatory coupling $(G_{c}\approx 0)$ the
behavior of the ENs is nearly independent and chaotic. As soon as the
strength of excitatory coupling is increased most of
the hyperpolarizing regions are in phase, and, eventually, for $G_{c}=100nS$
the oscillations are periodic and the bursts are synchronized. The
length of the bursts increases as we increase the coupling to $G_{c}>100nS$. 
For small inhibitory $G_{c}$ the oscillations are
chaotic but all the hyperpolarizing regions are in out-of-phase. These oscillations become
periodic as we approach $G_{c}=20nS$. For
the range $25nS<G_{c}<50nS$ the oscillations become chaotic again, but
out-of-phase bursting behavior is sustained. In the range
$50nS<G_{c}<150nS$ the oscillations are periodic with bursts out-of-phase,
and the bursting frequency decreases as we increase the
coupling. For $G_{c}>150nS$ chaotic spiking-bursting out-of-phase
behavior is observed, and the average burst length increases as we
increase the coupling up to $G_{c}=500nS$.}}
\label{mapchem}
\end{figure}

We have observed the behavior of two four dimensional ENs coupled with {\em %
identical} chemical synapses. Two electrical versions of chemical synapses
were built with identical parameters and then used to coupled two four
dimensional ENs. In the equations, Equation (\ref{chemsynapse}), we
integrate to represent the chemical synapse, all parameters were set equal
in the two connecting circuits. We then varied $G_c$ in each chemical
synapse over the range $0 \le G_c \le 200\,nS$ for an excitatory synapse,
namely $x_{\mbox{rev}}=-20mV$, and over the range $0 \le G_c \le 500\,nS$
for an inhibitory synapse, namely, $x_{\mbox{rev}}=-80mV$. The other
parameters were held fixed at $\tau _s=10ms , x_{\mbox{th}}=-50mV$, and $x_{%
\mbox{slope}}=10mV$. In Figure~\ref{mapchem} we collect the statistical
results, expressed in our usual quantities $\sigma_N(G_c)$ and $\Delta_N(G_c)
$, for both excitatory and inhibitory synaptic connections. Negative
values of $G_c$ represent inhibitory connections. This, perhaps apparently
peculiar, method of presentation allows us to see immediately the
relationship between excitatory and inhibitory interconnections. As earlier
with electrical couplings we provide a verbal description of each region of
behavior over the whole range of $G_c$. We show examples of the time series
for the membrane potential $x$ of the two neurons in Figure~\ref{exchem} and
Figure~\ref{inchem}.

\subsubsection{Excitatory Chemical Synapses}

\begin{figure}[ht!]
\centerline{
\epsfig{file=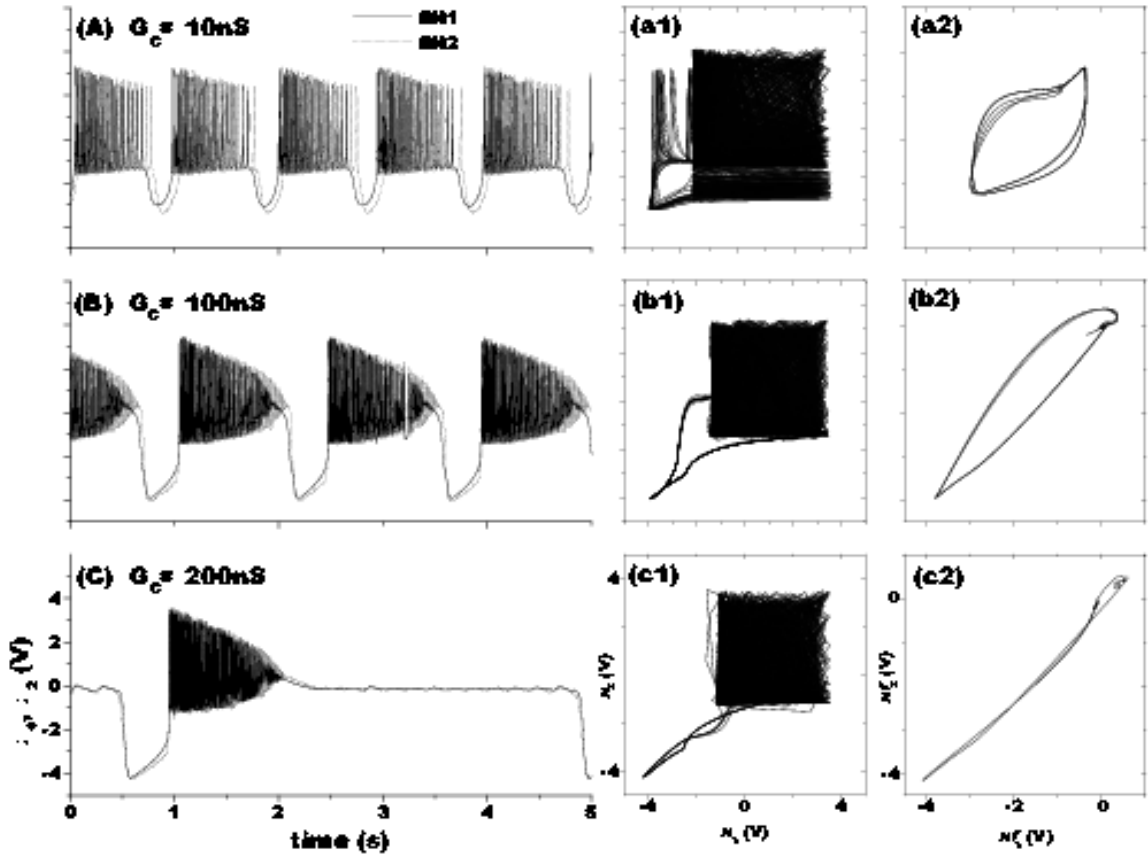,width=16cm}
}
\caption{\small Excitatory chemical coupling of two chaotic
ENs. Characteristic time series of the membrane potentials 
$x_1(t),x_2(t)$ (figure labels A to C). Phase 
portraits $x_{2}(t)$ vs. $x_{1}(t)$ (figure labels a1 to c1).
Phase portraits after 20Hz low pass filtering $x^{f}_{2}(t)$
vs. $x^{f}_{1}(t)$ (figure labels a2 to c2). (A) $G_{C}=10nS$--chaotic
but nearly synchronized bursting behavior. (B) $G_{C}=100nS$--periodic 
and synchronized bursting activity of the ENs. (C)
$G_{C}=200$ --periodic and synchronized activity with long bursts and
spikes vanishing before the end of the bursts.}
\label{exchem}
\end{figure}

When coupled with implementations of excitatory chemical synapses the ENs
displayed the following behaviors:

\begin{itemize}
\item  When $G_c \approx 0nS$ the two neurons are uncoupled and display
independent chaotic oscillations as shown in Figure~\ref{free}.

\item  For positive coupling $0<G_c<100nS$ a transition from the chaotic
behavior to regular spiking/bursting is observed. For small coupling the
independent chaotic spiking/bursting activity of the uncoupled neurons is
replaced by a behavior in which most of the bursts are synchronized, but the
oscillations are still chaotic as shown in Figure~\ref{exchem}(A) for $%
G_{c}=10nS$. As $G_{c}$ is increased all the bursts become synchronized, and
the activity becomes periodic as shown in Figure~\ref{exchem}(B) for $%
G_{c}=100nS$.

\item  For $G_c> 100nS$ the bursts remain synchronized and get longer, but
there are no longer any spikes during the bursts as shown in Figure~\ref
{exchem}(C) for $G_{c}=200nS$.
\end{itemize}

\subsubsection{Inhibitory Chemical Synapses}

Finally we report on our experiments with an electronic version of an
inhibitory chemical synapse. This inhibitory synaptic coupling occurs in the
lobster pyloric CPG as well as many other CPGs, and we have suggested~\cite
{neural} that inhibitory chemical coupling will lead to regularization of
the chaotic oscillations of the individual neurons.

\begin{figure}[ht!]
\centerline{
\epsfig{file=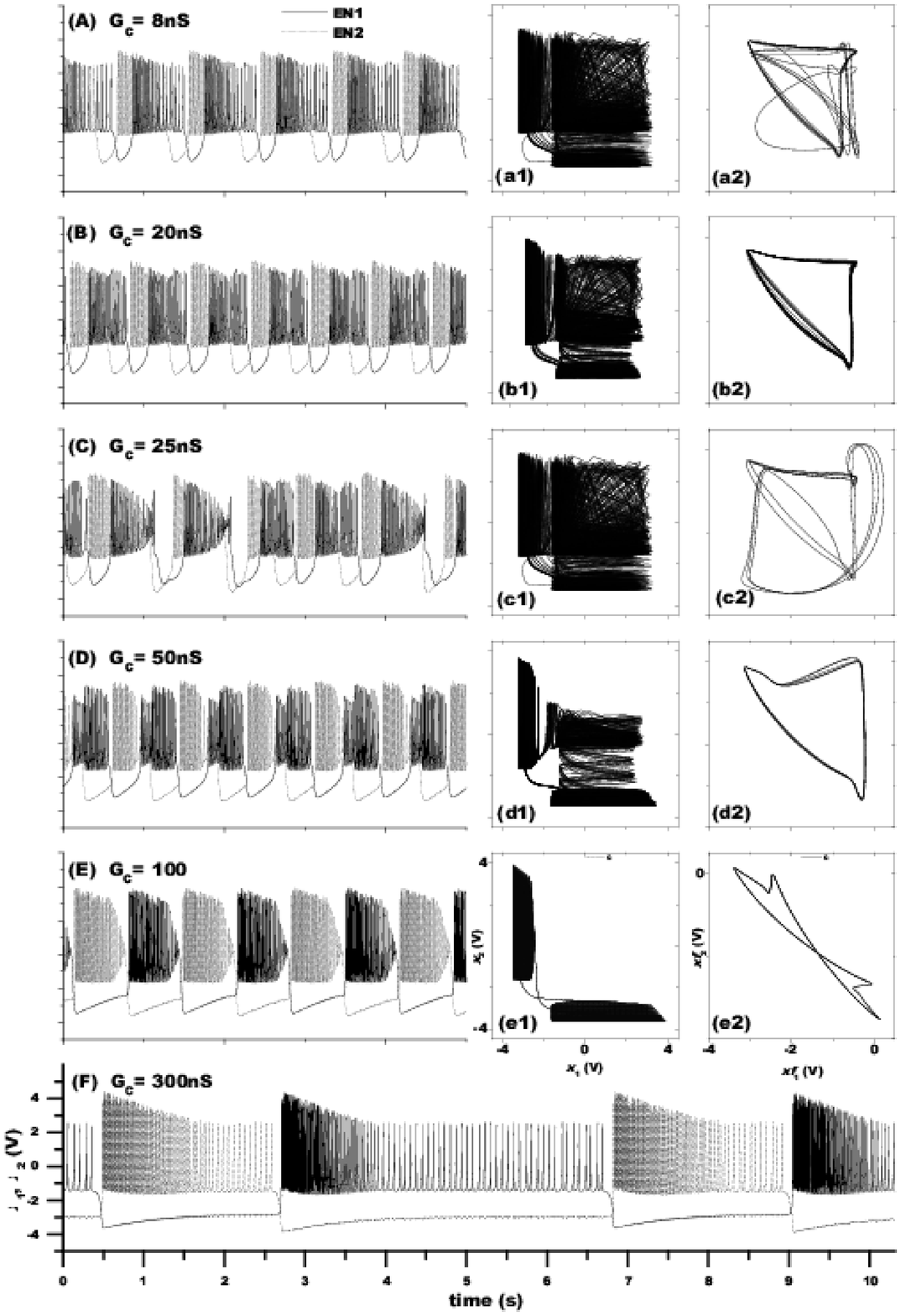,width=16cm}
}
\caption{\small Inhibitory chemical coupling of two chaotic
ENs. Characteristic time series of the membrane potentials $x_1(t),x_2(t)$ (figure labels A to F). Phase portraits $x_{2}(t)$ vs. $x_{1}(t)$ (figure labels a1 to f1).
Phase portraits after 20Hz low-pass filtering $x^{f}_{2}$
vs. $x^{f}_{1}$ (figure labels a2 to f). (A) $G_{C}=8nS$--chaotic
oscillations with all hyperpolarizing regions out-of-phase. (B)
$G_{C}=20nS$--periodic pattern with hyperpolarizing regions out-of-phase and some burst superposition. (C) $G_{C}=25nS$--chaotic
oscillations. (D) $G_{C}=50nS$--periodic out-of-phase bursting behavior
with some burst superposition. (E) $G_{C}=100nS$--periodic out-of-phase
spiking-bursting behavior. (F) $G_{C}=300nS$--chaotic out-of-phase
spiking-bursting pattern. The time scale used in the time series plot
is the same for all examples.}
\label{inchem}
\end{figure}

\begin{itemize}
\item  For small $G_{c}$ the oscillations are still chaotic, but all of the
hyperpolarizing regions of the membrane voltages are in anti-phase as shown
in Figure~\ref{inchem}(A) for $G_{c}=8nS$.

\item  When $G_{C} \approx 20nS$ the oscillations become periodic, and all
the hyperpolarizing regions are in out-of-phase as shown in Figure~\ref
{inchem}(B).

\item  For $25nS \leq G_{c}<50nS$ the out-of-phase behavior of the
hyperpolarizing regions remains, but the oscillations are chaotic again as
shown in Figure~\ref{inchem}(C) for $G_{c}=25nS$.

\item  For $50nS\leq G_{c}<150nS$ the oscillations regularize again, and the
behavior is periodic with out-of-phase bursting as shown in Figure~\ref
{inchem}(D) for $G_{c}=50nS$ and in Figure~\ref{inchem}(E) for $G_{c}=100nS$.

\item  For $G_{c}>150nS$ the oscillations are chaotic and long out-of-phase
bursts are observed as shown in Figure~\ref{inchem}(F) for $G_{c}=300nS$.
\end{itemize}

The only experiments we know which relate to these observations on two
chemically coupled ENs are not a precise match, but bear
noting. R. Elson~ \cite{rob} has isolated a pair of LP and PD neurons
from the lobster Pyloric CPG; these have mutual inhibitory
coupling. Elson varied the strength of the chemical coupling using
neuromodulators and making measurements at four values of $G_c$ over a
nominal rage of 20 nS to 60 nS. He observed only the behavior reported
in the penultimate item of our experiments on inhibitory
coupling. Unfortunately, control of the identity of the mutual
inhibitory couplings was not possible, nor was it possible for us to
directly compare the calibration of Elson's indication of the
magnitude of $G_c$ with our own choices in using ENs. To date then, we
have no direct laboratory evidence on synchronization of biological
neurons mutually coupled with chemical synapses. This is in contrast
to our observations on electrically coupled biological
neurons~\cite{prl}. This represents an interesting opportunity for biological experiments 
which may be directly compared to our results using ENs.

\subsection{Coupling Between Electronic and Living Neurons}

We have previously reported experiments on replacing the AB neuron from the
Pyloric CPG in its interaction with an isolated pair of PD neurons with a 
{\bf three dimensional} EN~\cite{Szu00}. For completeness in light of
the work reported in this paper, we carried out an experiment in which one
of our four dimensional neurons was coupled bidirectionally to one of the PD
neurons in the AB/PD pacemaker group of the Pyloric CPG. The full
description of the methods used in the biological preparation will appear
elsewhere~\cite{Szu001}, but here we quite briefly summarize those points
important to the main thrusts of this article.

These experiments were carried out on one of the two pyloric dilator (PD)
neurons from the pyloric central pattern generator (CPG) of the lobster
stomatogastric ganglion (STG)~\cite{Har92}. The STG of the California spiny
lobster, {\it Panulirus interruptus,} was removed using standard procedures
and pinned out in a dish lined with silicone elastomer and filled with
normal lobster saline. The STG was isolated from its associated anterior
ganglia, which provide activating inputs, by cutting the stomatogastric
nerve. Two glass microelectrodes were inserted in the soma of the PD neuron:
one for intracellular voltage recording and another one for current
injection. The voltage signals were digitized at 10000 samples/sec. The two
PD neurons remained coupled to each other and to the autonomously bursting
interneuron (AB) by their natural electrical synapses, but were isolated
from the rest of the CPG by blocking chemical input synapses with picrotoxin 
$(7.5\mu M)$. The artificial electrical coupling was provided by injecting
in the EN and in the PD opposite currents. More details of the experimental
setup can be found in~\cite{Szu00}. The membrane voltage of the EN was
reshaped to make its amplitude ratio in spiking/bursting mode, its total
amplitude and its voltage offset similar to those of the PD neuron. Only
electrical coupling, positive and negative, is reported here.

\begin{figure}[ht!]
\centerline{
\epsfig{file=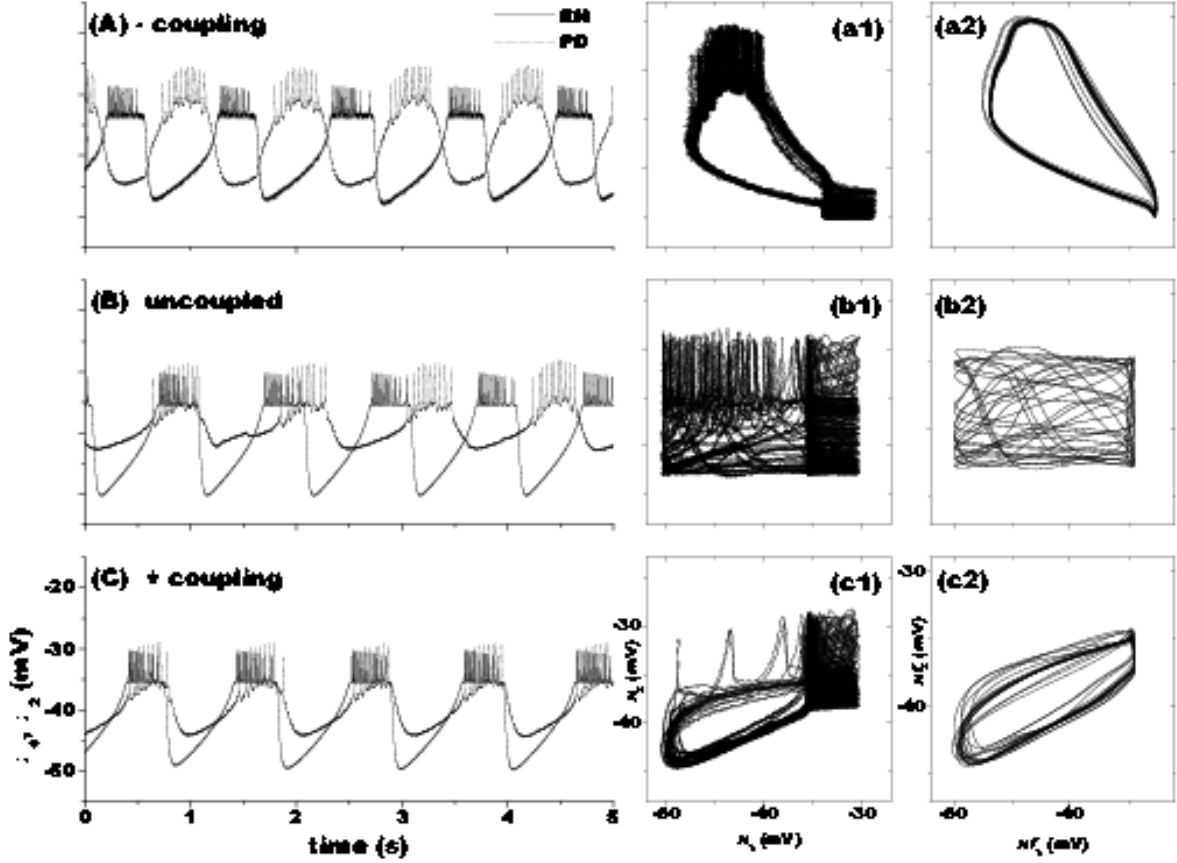,width=16cm}
}
\caption{\small Electrical coupling of an EN and a living pyloric
dilator (PD) neuron from the stomatogastric ganglion (STG) of the
lobster \textit{Panulirus interruptus}. Characteristic time series of
the membrane potentials $x_1(t),x_2(t)$ (figure labels A to C). Phase portraits of $x_{2}(t)$
vs. $x_{1}(t)$ (figure labels a1 to c1). Phase portraits after 20Hz
low-pass filtering $x^{f}_{2}(t)$ vs. $x^{f}_{1}(t)$ (figure labels a2 to c2). 
(A) Negative coupling--out-of-phase bursting activity. (B)
Uncoupled neurons. (C) Positive electrical coupling--synchronized
bursting behavior}
\label{hybrid}
\end{figure}

We connected the neurons with the analog electrical synapse and observed
their spiking-bursting behavior as shown in Figure~\ref{hybrid}. When
uncoupled, the neurons had independent spiking/bursting activity as shown in
Figure~\ref{hybrid}(B). For large enough negative coupling the neurons are
synchronized and fire out-of-phase as shown in Figure~\ref{hybrid}(A). For
positive coupling the neurons show synchronized bursting activity as shown
in Figure~\ref{hybrid}(C). For this value of $G_E$ the bursts are
synchronized but not the spikes.

This result is in agreement with the experiments made with a pair of
electrically coupled ENs, as we discussed above, as well as for a pair of
living STG neurons~\cite{prl}.

\section{Discussion}

The ENs described in this paper are simple analog circuits which
integrate four dimensional differential equations representing fast
and slow subcellular processes that give rise to the characteristic
spiking and spiking-bursting behavior of CPG neurons. The single
neurons can be easily set into a chaotic regime that reproduces the
irregular firing patterns observed in biological neurons when isolated
from the rest of the CPG. This study comprises: (a) two electrically
coupled ENs and (b) two ENs connected with excitatory and inhibitory
chemical synapses. These two types of connections exist in almost all
known CPGs. The range of observations summarized in
Figures~\ref{mapel} and~\ref{mapchem} shows the rich behavior and
complexity of these minimal network configurations. It indicates how
small changes in the coupling conductance can drive the cells into
completely different regimes. In particular, some of our experiments
predict the appearance of chaotic out-of-phase synchronization for
different coupling configurations. These results are displayed in
Figures~\ref{negel}C and~\ref{inchem}F. In general, the
experiments with the ENs contribute directly to our understanding of
the origin of regularization of individually chaotic neurons through
cooperative activity.

How complicated should one require a model neuron to be? In our view the
answer depends on the neural function one wishes to represent. The analysis
of the electrical activity of isolated neurons from the lobster Pyloric CPG
indicates that the number of active degrees of freedom is not very large,
ranging from three to five in various environments, and this suggests a very
simple representation in terms of dynamical equations. Our analysis~\cite
{martin} of much richer Hodgkin-Huxley models of these individual neural
oscillators also indicates that in the regime of biological operation, the
number of active degrees of freedom is equally small. On this basis we
developed the Hindmarsh-Rose type models of these neurons both in numerical
simulation and in analog electrical circuitry.

This paper has moved that inquiry about the complexity of representation for
the components of a biologically realistic neural network to another level.
Here we have investigated whether the simplified neural models, when coupled
together in small networks, but in biologically realistic manners, can
reproduce our observations biological neurons alone. The striking result of
the observations presented here, when the experimental setup matches that of
the biological networks, is that the observed behavior of the ENs also
matches. Further, using our ENs, we are able to make distinct predictions
about the behavior of biological or hybrid (biological and EN) networks in
settings not yet investigated.

Our experiments on coupled biological neurons and ENs provide
further ground for testing the validity of numerical and electronic models
of individual neural behavior as well as presenting interesting new examples
of coupled nonlinear oscillators. Hybrid circuits with biological and
electronic neurons coupled together are a powerful mechanism to understand
the modes of operation of CPGs. The hybrid system constitutes an easy way to
change the connectivity and global topology of the CPG. The roles of
intrinsic dynamics of the neurons and the synaptic properties of the network
in rhythm generation can be easily studied with these hybrid configurations~
\cite{Szu00}).

There are previous efforts studying electronic neurons alone and in
conjunction with biological neurons. An early example is the work of Yarom~
\cite{yarom} where a network of four oscillators, realized as an analog
circuit, was interfaced with an olivary neuron in a slice preparation. Yarom
studied the response of the olivary neuron when it received oscillating
electrical input from the network. There was no feedback from the biological
neuron to the network he constructed. Le Masson, et al~\cite{lemasson1} developed a digital version of a neuron
comprising a Hodgkin-Huxley~\cite{HH} (HH) model of various pyloric CPG
neurons with three compartments and eight different ion channels which ran
on a DSP board located on the bus of a personal computer. They connected
this model into a variety of different configurations of subcircuits of the
pyloric CPG replacing at various times the LP, a PD or a PY neuron. Using
this `hybrid' setup they verified that many aspects of the pyloric rhythm
are accurately reproduced when their DSP based neuron replaces one of the
biological neurons in their system. In subsequent work~\cite
{lemasson2,lemasson3}, this group has developed VLSI devices for integrating
the HH models and has utilized them in mixed circuits (ENs and biological
neurons), replacing the DSP version of the conductance models in their
biological preparations. The complexity of these ENs has not been needed in
our modeling nor in the further experiments on their interaction with each
other as reported here. We have not found any reports in the literature on
the mutual interaction of these analog VLSI neural circuits.

There are two interesting directions to which the results reported here may
point:

\begin{itemize}
\item  (1) biologically realistic neural networks of much greater size than
the elementary ones investigated here may be efficiently investigated
numerically or in analog circuitry using the realistic, but simple HR type
models. The integration of the model equations is no challenge to easily
available computing power and large networks should be amenable to
investigation and analysis.

\item  (2) The networks investigated here are subcircuits of a biological
circuit of about fifteen neurons which has the functional role of a control
system: commands are presented from other ganglia of the lobster and this
Pyloric circuit must express voltage activity to the muscles to operate a
pump for shredded food passing from the stomach to the digestive system.
Many other functions are asked of biological neural networks. Using the full
richness of HH models for the component neurons may seem attractive at one
level, but the results presented here suggest that many interesting
questions may be asked of those networks using the simplified
component neurons studied here.
\end{itemize}

\section*{Acknowledgements}

R.D. Pinto was supported by the Brazilian Agency Funda\c{c}\~{a}o de Amparo
\`a Pesquisa do Estado de S\~{a}o Paulo - FAPESP. PV acknowledges support
from MEC. Partial support for this work came from the U.S Department of
Energy, Office of Science, under grants DE-FG03-90ER14138 and
DE-FG03-96ER14592. We also acknowledge the many conversations we have had
with Ramon Huerta, Rob Elson and Allen Selverston on the dynamics of CPG
neurons.

\clearpage
 
\clearpage

\clearpage

\clearpage

\clearpage

\clearpage

\clearpage

\clearpage

\clearpage

\end{document}